\documentclass{PoS}

\def\bea{\begin{eqnarray}}
\def\eea{\end{eqnarray}}
\def\be{\begin{equation}}
\def\ee{\end{equation}}

\def\eps{\varepsilon}

\def\ms{M_\odot}

\title{Constraints on the quark matter equation of state from astrophysical observations}

\ShortTitle{Constraints on the quark matter equation of state from astrophysical observations}

\author{\speaker{G. F. Burgio}\\
        INFN Sezione di Catania, via S. Sofia 64, I-95123 Catania, Italy\\
        E-mail: \email{fiorella.burgio@ct.infn.it}}

\author{H. Chen\\
       Physics Department, China University of Geoscience, Wuhan 430074, China\\
        E-mail: \email{huanchen.phy@gmail.com}}

\author{H.-J. Schulze\\
       INFN Sezione di Catania, via S. Sofia 64, I-95123 Catania, Italy\\
        E-mail: \email{schulze@ct.infn.it}}

\author{G. Taranto \\
       Dipartimento di Fisica e Astronomia, Universita' di Catania, and INFN Sezione di Catania,
        via S. Sofia 64, I-95123 Catania, Italy\\
        E-mail: \email{tarantogabriele@gmail.com}}

\abstract{
We calculate the structure of neutron star interiors comprising both
the hadronic and the quark phases. 
For the hadronic sector we employ a microscopic equation
of state involving nucleons and hyperons derived within the
Brueckner-Hartree-Fock many-body theory with realistic two-body and
three-body forces. For the description of quark matter, we use several different models,
e.g. the MIT bag, the Nambu--Jona-Lasinio (NJL), the Color Dielectric (CDM), the Field 
Correlator method (FCM), and one 
based on the Dyson-Schwinger model (DSM). We find that a two solar mass hybrid star 
is possible only if the nucleonic EOS is stiff enough.}

\FullConference{Xth Quark Confinement and the Hadron Spectrum,\\
		October 8-12, 2012\\
		TUM Campus Garching, Munich, Germany}

\begin{document}

The possible appearance of quark matter (QM)
in the interior of massive neutron stars
(NS) is one of the main issues in the physics of these compact objects.
Calculations of NS structure, based on a microscopic nucleonic
equation of state (EOS), indicate that for the heaviest NS, close
to the maximum mass (about two solar masses), the central particle density
reaches values larger than $1/\rm fm^{3}$.
In this density range the nucleons start to loose their identity,
and quark degrees of freedom play a role. 

The value of the maximum mass of a NS is probably one of the physical
quantities that are most sensitive to the presence of QM, and
the recent claim of discovery of a two solar mass NS  \cite{heavy}
has stimulated the interest in this issue.
Unfortunately, while the microscopic theory of the nucleonic EOS has
reached a high degree of sophistication \cite{rep}, the QM EOS is poorly
known at zero temperature and at the high baryonic density appropriate for NS.
One has, therefore, to rely on models of QM, which contain
a high degree of arbitrariness.
At present, the best one can do is to compare the predictions of
different quark models and to estimate the uncertainty of the results
for the NS matter as well as for the NS structure and mass.

In this paper we will discuss a set of different quark models in combination with a definite 
baryonic EOS, which has been developed within the Brueckner-Hartree-Fock (BHF)
many-body approach of nuclear matter, comprising nucleons and also hyperons \cite{book}.
In particular, in section \ref{s:bhf} we review the derivation of the baryonic
EOS in the BHF approach, whereas Section \ref{s:qm} is devoted to the relevant features of the hadron quark phase transition for the several QM EOS discussed. 
In section \ref{s:ns} we present the results regarding NS structure,
combining the baryonic and QM EOS for beta-stable nuclear matter, and conclusions are 
drawn.

\section{EOS of nuclear matter within Brueckner theory}
\label{s:bhf}

The EOS constructed for the hadronic phase at $T=0$ is based on the
non-relativistic Brueckner-Bethe-Goldstone (BBG) many-body theory \cite{book},
which is a linked cluster
expansion of the energy per nucleon of nuclear matter, well convergent
and accurate enough in the density range relevant for neutron stars.
In this approach the essential ingredient is the two-body scattering matrix
$G$, which,
along with the single-particle potential $U$, satisfies the
self-consistent equations

\begin{eqnarray}
G(\rho;\omega)& = & V  + V \sum_{k_a k_b} {{|k_a k_b\rangle  Q  \langle k_a k_b|}
  \over {\omega - e(k_a) - e(k_b) }} G(\rho;\omega), \\
U(k;\rho) &= &\sum _{k'\leq k_F} \langle k k'|G(\rho; e(k)+e(k'))|k k'\rangle_a,
\end{eqnarray}
where $V$ is the bare nucleon-nucleon (NN) interaction,
$\rho$ is the nucleon number density,
$\omega$  is the  starting energy,
$|k_a k_b\rangle Q \langle k_a k_b|$ is the Pauli operator,
$e(k) = e(k;\rho) = {{\hbar^2}\over {2m}}k^2 + U(k;\rho)$
is the single particle energy,
and the subscript ``{\it a}'' indicates antisymmetrization of the
matrix element.
In the BHF approximation the energy per nucleon is
\begin{eqnarray}
&&{E \over{A}}(\rho)  =
          {{3}\over{5}}{{\hbar^2~k_F^2}\over {2m}} + D_{\rm 2}\, , \\
&&D_{\rm 2} = {{1}\over{2A}}
\sum_{k,k'\leq k_F} \langle k k'|G(\rho; e(k)+e(k'))|k k'\rangle_a.
\end{eqnarray}
The inclusion of nuclear three-body forces (TBF) is crucial in order to
reproduce the correct saturation point of symmetric nuclear matter \cite{bbb,akma}.
The present theoretical status of microscopically derived TBF is quite
rudimentary.
Recent results \cite{zhli,zuotbf} have shown that both two-body and three-body 
forces should be based on the same theoretical
footing and use the same microscopical parameters in their construction.
Results shown here were obtained with the Argonne $V_{18}$ (V18) \cite{v18}
or the Bonn B (BOB) \cite{bob} potentials, and compared also with the widely
used phenomenological Urbana-type (UIX) TBF \cite{uix}
(in combination with the V18 potential).
It should be stressed that in our approach the TBF is reduced to
a density-dependent two-body force by averaging over the position of the
third particle, assuming that the probability of having two particles at a given distance
is given by the two-body correlation function determined self-consistently.

In the past years, the BHF approach has been extended with the inclusion of hyperons \cite{sch98,vi00,mmy}, which may appear at baryon density of about 2 to 3 times normal nuclear matter density.
The inclusion of hyperons produces an EOS which turns out to be much softer than the purely nucleonic case, with dramatic consequences for the structure of the NS, i.e., the value of the maximum mass is smaller than the canonical value 1.44 $M_\odot$.
The inclusion of further theoretical ingredients,
such as hyperon-hyperon potentials \cite{vi00}
and/or three-body forces involving hyperons,
could alter the baryonic EOS, but unfortunately they are essentially unknown. 
Another possibility that is able to produce larger maximum masses,
is the appearance of a transition to QM inside the star.
This scenario will be illustrated below.

\begin{figure}
\centering
\includegraphics[scale=0.8,clip]{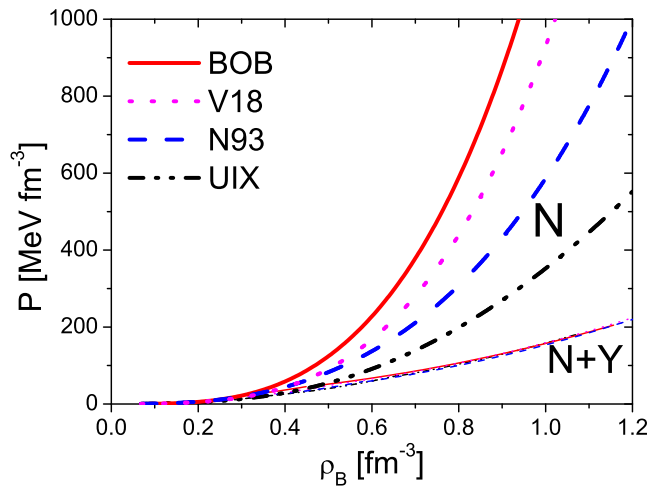}
\caption{
Pressure vs.~the baryon number density of hadronic NS matter.
Thick curves show results for purely nucleonic matter,
whereas thin curves include hyperons.
\label{f:eos}}
\end{figure}

Starting from the EOS for symmetric and pure neutron matter,
and assuming stellar matter
composed of neutrons, protons, and leptons \cite{bbb}, the
EOS for the beta equilibrated matter can be obtained in the usual standard way
\cite{bbb,shapiro,gle}:
The Brueckner calculation yields the energy density of
baryon/lepton matter as a function of the different partial densities,
$\eps(\rho_n,\rho_p,\rho_e,\rho_\mu)$.
The various chemical potentials for the species $i=n,p,e,\mu$ can then be computed straightforwardly,
\be
 \mu_i = {\partial \eps \over \partial \rho_i} \:,
\ee
and the equations for beta-equilibrium,
\be
\mu_i = b_i \mu_n - q_i \mu_e \:,
\ee
($b_i$ and $q_i$ denoting baryon number and charge of species $i$)
and charge neutrality,
\be
 \sum_i \rho_i q_i = 0 \:,
\ee
allow one to determine the equilibrium composition $\{\rho_i(\rho)\}$
at given baryon density $\rho$ and finally the EOS,
\be
 P(\rho) = \rho^2 {d\over d\rho}
 {\eps(\{\rho_i(\rho)\})\over \rho}
 = \rho {d\eps \over d\rho} - \eps
 = \rho \mu_n - \eps \:.
\ee
In Fig.~\ref{f:eos} we compare the EOS obtained in the BHF framework when only
nucleons and leptons are present (thick lines),
and the corresponding ones with hyperons included (thin lines).
Calculations have been performed with different choices of the NN potentials,
i.e., the Bonn B, the Argonne V18, and the Nijmegen N93,
all supplemented by a compatible microscopic TBF \cite{zhli}.
For completeness, we also show results obtained with the Argonne V18 potential
together with the phenomenological Urbana IX as TBF.

We notice a strong dependence on both the chosen NN potential,
and on the adopted TBF, the microscopic one being more
repulsive than the phenomenological force. 
The presence of hyperons decreases strongly the pressure,
and the resulting EOS turns out to be almost independent of the
adopted NN potential, due to the interplay between the stiffness of the
nucleonic EOS and the threshold density of hyperons \cite{zhli}.
The softening of the EOS has serious consequences for the structure of NS,
leading to a maximum mass of less than 1.4 solar masses \cite{zhli,mmy},
which is below the observed pulsar masses \cite{obs}.

\section{Quark Phase}
\label{s:qm}

The properties of cold nuclear matter at large densities, i.e.,
its EOS and the location of the phase transition to deconfined QM,
remain poorly known.
The difficulty in performing first-principle calculations in such systems
can be traced back to the complicated nonlinear and nonperturbative nature of
quantum chromodynamics (QCD).
Therefore one can presently only resort to more or less phenomenological models
for describing QM. 

\begin{figure}
\centering
\includegraphics[scale=0.5,clip]{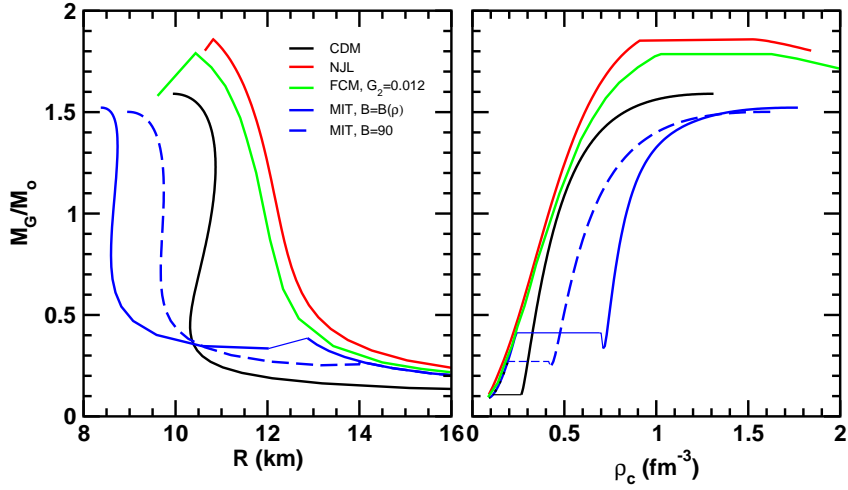}
\caption{
Gravitational NS mass vs.~the radius (left panel)
and the central baryon density (right panel) for different EOS
of quark matter. See text for details.
\label{fig:f1}}
\end{figure}

One of the most widely used approach is the MIT bag model \cite{chodos}. In the calculations
shown below, we assumed massless $u$ and $d$ quarks,
$s$ quarks with a current mass of $m_s=150$ MeV, and either a fixed bag constant
$B = 90\;\rm MeV\,fm^{-3}$, or a density-dependent bag parameter,
\bea
 B(\rho) = B_\infty
 + (B_0 - B_\infty) \exp\!\Big[-\beta \Big( \frac{\rho}{\rho_0} \Big )^2\Big ]
\label{e:brho}
\eea
with $B_\infty = 50\;\rm MeV\,fm^{-3}$,
$B_0=400\;\rm  MeV\,fm^{-3}$,
and $\beta=0.17$.
This approach has been proposed in \cite{nsquark}, and it
allows the symmetric nuclear matter to be in the pure hadronic phase at
low densities, and in the quark phase at large densities,
while the transition density is taken as a parameter. We find  that the phase
transition takes place at baryon density of the order of 2--3$\rho_0$,
and that the mixed phase contains a small fraction of hyperons.

The phase transition from the hadronic phase to the pure quark matter phase
is usually performed through the Maxwell construction, or the more sophisticated Gibbs
construction \cite{gle}. This is widely used for the study of neutron star structure, several numerical details 
can be found in many references \cite{gle}, and will not be repeated here.
The important point to stress is that the features of the phase transition, such as its onset and the
density interval over which it extends, depend crucially on the models used for
describing the hadron and the quark phase. 
For example, we found that the phase transition constructed with the CDM \cite{cdm} model is quite different from the one obtained using the MIT bag model.
In the CDM, the onset of the coexistence region occurs at very low baryonic density. This implies a large difference in the structure of neutron stars.
In fact, whereas stars built with the CDM have at most a mixed phase
at low density and a pure quark core at higher density, the ones obtained
with the MIT bag model contain a hadron phase, followed by a mixed phase
and a pure quark interior. The scenario is again different  within the Nambu-Jona-Lasinio model \cite{njl}, where at most a mixed phase is present, but no pure quark phase.
It has also been found that the phase transition from hadronic to QM
occurs at high values of the baryon chemical potential when the Dyson-Schwinger model
is used to describe the quark phase. 
In some extreme cases, for particular choices of the parameters, no phase transition at all is possible.
In fact, the Dyson-Schwinger EOS \cite{ds} is generally stiffer than the hadronic one,
and the value of the transition density is high. We also found that with the DSM no phase transition exists if the hadronic phase contains hyperons, just like the phase transition with NJL model.

It is worthwhile to mention that neutron stars observations can help to
determine free parameters, which could be present in some quark matter models.
This is the case of the QM EOS based on the Field Correlator method \cite{fcm},
which depends crucially on the value of the gluon condensate $G_2$. It turns out that using 
value of $G_2\simeq 0.006-0.007~ \rm GeV^4$, which gives a critical temperature 
$\rm T_c\simeq 170~ MeV$, produces maximum masses which are only marginally
consistent with the observational limit, while larger masses are possible if the value of the 
gluon condensate is increased. 
Also in this case, the phase transition only takes place if no
hyperons are present in the hadronic phase.

\section{Neutron star structure}
\label{s:ns}

\begin{figure}[t]
\centering
\includegraphics[scale=0.35,clip]{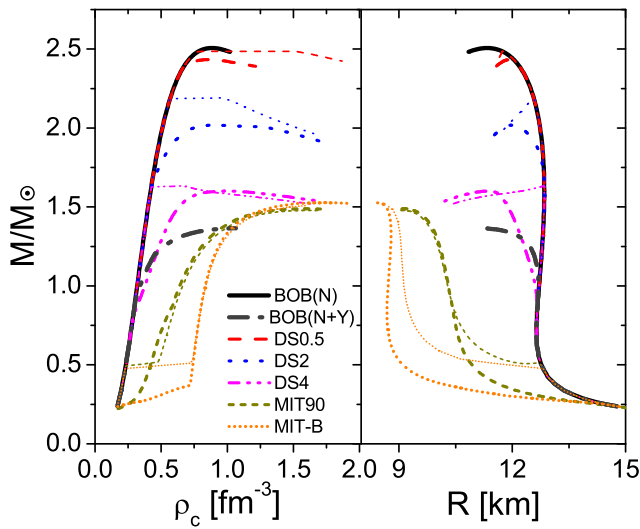}
\caption{
Gravitational NS mass vs.~the radius (right panel)
and the central baryon density (left panel) for different EOS
employing the BOB hadronic model.
\label{fig:f2}}
\end{figure}

We assume that a NS is a spherically symmetric distribution of
mass in hydrostatic equilibrium.
The equilibrium configurations are obtained
by solving the Tolman-Oppenheimer-Volkoff (TOV) equations \cite{shapiro} for
the pressure $P$ and the enclosed mass $m$,
\begin{eqnarray}
  {dP\over dr} &=& -{ G m \eps \over r^2 }
  {  \left( 1 + {P / \eps} \right)
  \left( 1 + {4\pi r^3 P / m} \right)
  \over
  1 - {2G m/ r} } \:,
\\
  {dm \over dr} &=& 4 \pi r^2 \eps \:,
\end{eqnarray}
being $G$ the gravitational constant.
Starting with a central mass density $\eps(r=0) \equiv \eps_c$,
we integrate out until the density on the surface equals the one of iron.
This gives the stellar radius $R$ and the gravitational mass is then
\be
 M_G \equiv m(R) = 4\pi \int_0^R dr\; r^2 \eps(r) \:.
\ee
We have used as input the EOS discussed above, and the results are plotted in Figs.~\ref{fig:f1} and \ref{fig:f2},
where we display the gravitational mass $M_G$
(in units of the solar mass $M_\odot= \rm 2 \times 10^{33}g $)
as a function of the radius $R$ and central baryon density $\rho_c$.

Calculations displayed in Fig.~\ref{fig:f1} for neutron star matter are obtained in the 
BHF theoretical framework with the V18 nucleon-nucleon potential in combination with
i) the CDM (solid black curve), and ii) the MIT bag model (blue curves) for quark matter. 
Due to the use of the Maxwell construction, the curves are not continuous \cite{gle}:
for very small central densities (large radii, small masses)  the stars are purely hadronic.
We observe that the values of the maximum mass depend only slightly 
on the EOS chosen for describing quark matter, and lie between 1.5 and 1.6 solar masses.
A clear difference between the two models exists as far as the radius 
is concerned. Hybrid stars built with the CDM are characterized by a larger radius
and a smaller central density, whereas hybrid stars constructed with the MIT bag model are more compact. Further calculations have been performed using the Paris potential as nucleon-nucleon  interaction for the hadronic phase, and the NJL model for the quark phase (red curve). In this case
the onset of pure quark matter leads to an instability, as well as in the FCM (green curve).
In both cases the phase transition takes place only if no hyperons are present in the 
hadronic phase. Unfortunately, for all the cases discussed above, the value of the maximum 
mass lies below the mass observed for the pulsar PSR J1614-2230, $\ms = 1.97 \pm 0.04$ \cite{heavy}.
Such a high value puts severe constraints on the EOS, and in particular it demands
an additional repulsion in the QM EOS. 

In Fig.~\ref{fig:f2} we show results obtained in the BHF framework using the Bonn B nucleon-nucleon potential, which produces the stiffest EOS, as shown in Fig.~\ref{f:eos}.
This yields a very high maximum NS mass, $\approx2.50\;\ms$, with only nucleons (solid black curve),
and $1.37\;\ms$ including hyperons. Using the DSM for the quark phase, we found 
that the maximum mass of hybrid stars is only a little lower than 2.5 $\ms$ with $\alpha=0.5$,
and decreases to about 2 $\ms$ with $\alpha=2$, being $\alpha$ a model parameter 
which controls the rate of approaching asymptotic freedom.
With increasing $\alpha$ we can obtain a smooth change from the pure hadronic NS to the
results with the MIT bag model. Moreover, 
no phase transition can occur and no hybrid star can exist if hyperons are introduced. 
If hyperons are excluded, the phase transition from nucleon matter to QM takes place
at rather large baryon density, and 
the onset of the phase transition is determined in this case by the parameter $\alpha$.

The possible effects of the hadron-quark phase transition are very
different with the MIT bag model and the DSM: In the case of the MIT
model, the phase transition begins at very low baryon density and
thus effectively impedes the appearance of hyperons.
Consequently the resulting maximum mass of the MIT hybrid star is
$1.5\;\ms$, lower than the value of the nucleonic star, but higher
than that of the hyperon star given before. A clear difference between the two models exists as far as the radius
is concerned. Hybrid stars built with the DSM are characterized by
a larger radius and a smaller central density,
whereas hybrid stars constructed with the MIT bag model are more compact.

In conclusion, a hybrid star with 2 $\ms$ is only allowed if the nucleonic EOS is stiff enough,
and the hadron-quark phase transition takes place without hyperons in the hadronic phase.

\end{document}